\documentclass[11pt]{article}
\usepackage{amssymb,amsmath,fullpage,graphicx}

\def\x{\mathbf x}

\makeatletter \numberwithin{equation}{section}

\begin{document}
\title{Classification of solitary wave bifurcations in generalized nonlinear Schr\"odinger equations}
\author{Jianke Yang \\
Department of Mathematics and Statistics \\ University of Vermont \\
Burlington, VT 05401, USA}
\date{ }
\maketitle

\begin{abstract}
Bifurcations of solitary waves are classified for the generalized
nonlinear Schr\"odinger equations with arbitrary nonlinearities and
external potentials in arbitrary spatial dimensions. Analytical
conditions are derived for three major types of solitary wave
bifurcations, namely saddle-node bifurcations, pitchfork
bifurcations and transcritical bifurcations. Shapes of power
diagrams near these bifurcations are also obtained. It is shown that
for pitchfork and transcritical bifurcations, their power diagrams
look differently from their familiar solution-bifurcation diagrams.
Numerical examples for these three types of bifurcations are given
as well. Of these numerical examples, one shows a transcritical
bifurcation, which is the first report of transcritical bifurcations
in the generalized nonlinear Schr\"odinger equations. Another shows
a power loop phenomenon which contains several saddle-node
bifurcations, and a third example shows double pitchfork
bifurcations. These numerical examples are in good agreement with
the analytical results.

\end{abstract}

\section{Introduction}

Solitary waves are spatially localized and temporally stationary (or
steadily moving) solutions of nonlinear wave equations. Solitary
waves play an important role in the understanding of nonlinear wave
dynamics and thus have been heavily studied for a wide range of
nonlinear wave models arising in diverse physical disciplines
\cite{Kivshar_book,Yang_SIAM}. When the propagation constant of
solitary waves or physical parameters in the nonlinear wave
equations changes, bifurcations of solitary waves can occur. Indeed,
various solitary wave bifurcations in miscellaneous nonlinear wave
models have been reported. Examples include saddle-node bifurcations
(also called fold bifurcations)
\cite{Yang_SIAM,Champneys_1996,Yang_CPNLS_1997,Akylas_1997,Chen_2000,trans_saddle_node_4_wave,Panos_2005,Kapitula_2006,Burke_2007,Akylas_2012},
pitchfork bifurcations (sometimes called symmetry-breaking
bifurcations)
\cite{Akylas_2012,Akhmediev_pitchfork_1995,Weinstein_2004,Panos_2005_pitchfork,Weinstein_2008,Sacchetti_2009,Panos_2009,Kirr_2011,Malomed_pitchfork},
transcritical bifurcations \cite{trans_saddle_node_4_wave}, and so
on. Most of these reports on bifurcations are numerical. In the few
analytical studies, focus was on the quantitative prediction of
symmetry-breaking bifurcation points in the nonlinear Schr\"odinger
(NLS) equations with symmetric double-well potentials
\cite{Weinstein_2004,Weinstein_2008,Sacchetti_2009,Kirr_2011,Malomed_pitchfork}
and the prediction of saddle-node and pitchfork bifurcation points
in the NLS equations with periodic potentials \cite{Akylas_2012}. A
general treatment of these bifurcations and general analytical
conditions for their occurrences are still lacking at this time.

In this paper, we systematically classify solitary wave bifurcations
in the generalized NLS equations with arbitrary nonlinearities and
external potentials in arbitrary spatial dimensions. These
generalized NLS equations include the Gross-Pitaevskii equations in
Bose-Einstein condensates \cite{BEC1} and nonlinear
light-transmission equations in refractive-index-modulated optical
media \cite{Kivshar_book,Yang_SIAM} as special cases. For this large
class of wave equations, we derive sufficient analytical conditions
for three major types of solitary wave bifurcations, namely
saddle-node bifurcations, pitchfork bifurcations and transcritical
bifurcations. In addition, shapes of power diagrams near these
bifurcation points are also derived. We will show that the power
diagram near a saddle-node bifurcation is a horizontally oriented
parabola; the power diagram near a pitchfork bifurcation is an extra
power curve bifurcating out from a smooth power curve on one side of
the bifurcation point (like a slanted letter `y'); and the power
diagram near a transcritical bifurcation comprises two smooth curves
tangentially connected at the bifurcation point. These analytical
results are followed by various numerical examples. One example
shows a transcritical bifurcation, which is the first report of
transcritical bifurcations in the generalized NLS equations to the
author's best knowledge. Another example shows double pitchfork
bifurcations combined with saddle-node bifurcations, and a third
example shows a power loop phenomenon which contains a number of
saddle-node bifurdations. These numerical examples of bifurcations
are found to be in good agreement with our analytical results.

\section{Preliminaries}
We consider the generalized nonlinear Schr\"odinger (GNLS) equations
with arbitrary forms of nonlinearity and external potentials in any
spatial dimensions. These equations can be written as
\begin{equation}  \label{e:U}
iU_t+\nabla^2 U+F(|U|^2, \x)U=0,
\end{equation}
where $\nabla^2=\partial^2/\partial x_1^2+\partial^2/\partial
x_2^2+\cdots + \partial^2/\partial x_N^2$ is the Laplacian in the
$N$-dimensional space $\textbf{x}=(x_1, x_2, \cdots, x_N)$, and
$F(\cdot, \cdot)$ is a general real-valued function which includes
nonlinearity as well as external potentials. These GNLS equations
are physically important since they include the Gross-Pitaevskii
equations in Bose-Einstein condensates \cite{BEC1} and nonlinear
light-transmission equations in refractive-index-modulated optical
media \cite{Kivshar_book,Yang_SIAM} as special cases. Notice that
these GNLS equations are conservative and Hamiltonian.

For a large class of nonlinearities and potentials, these GNLS
equations admit stationary solitary waves
\begin{equation}  \label{e:Usoliton}
U(\x,t)=e^{i\mu t}u(\x),
\end{equation}
where $u(\x)$ is a real localized function in the square-integrable
functional space which satisfies the equation
\begin{equation}  \label{e:u}
\nabla^2u-\mu u+F(u^2, \x)u=0,
\end{equation}
and $\mu$ is a real-valued propagation constant. Examples of such
solitary waves can be found in numerous books and articles (see
\cite{Kivshar_book,Yang_SIAM} for instance). In these solitary
waves, $\mu$ is a free parameter, and $u(\x)$ depends continuously
on $\mu$. Under certain conditions, these solitary waves undergo
bifurcations at special values of $\mu$. Reported examples of
bifurcations in Eq. (\ref{e:u}) include saddle-node bifurcations
\cite{Yang_SIAM,Panos_2005,Kapitula_2006,Akylas_2012} and pitchfork
bifurcations
\cite{Akylas_2012,Weinstein_2004,Panos_2005_pitchfork,Weinstein_2008,Sacchetti_2009,Panos_2009,Kirr_2011,Malomed_pitchfork}.
Transcritical bifurcations in this equation have not been reported
yet (even though they have been found in other nonlinear wave models
\cite{trans_saddle_node_4_wave}).

For later analysis, we introduce the linearization operator of Eq.
(\ref{e:u}),
\begin{equation}
L_1=\nabla^2-\mu+\partial_u[F(u^2, \x)u],
\end{equation}
which is a self-adjoint linear Schr\"odinger operator. We also
introduce the standard inner product of functions,
\begin{equation} \label{def:inner_prod}
\langle f, g\rangle  = \int_{-\infty}^\infty f^*(\x) \hspace{0.05cm}
g(\x) \hspace{0.07cm} d \x.
\end{equation}
In addition, we define the power of a solitary wave $u(\x;\mu)$ as
\begin{equation}
P(\mu)=\langle u, u\rangle = \int_{-\infty}^\infty u^2(\x; \mu)
\hspace{0.07cm} d \x.
\end{equation}
This power function not only conveniently characterizes
solitary-wave families, but also plays an important role in the
stability of these waves \cite{Yang_SIAM}.

Our analysis of bifurcations starts with the basic observation that,
if a bifurcation occurs at $\mu=\mu_0$, by denoting the
corresponding solitary wave and the linearization operator as
\begin{equation}
u_0(\x)=u(\x; \mu_0), \quad L_{10}=L_1|_{\mu=\mu_0,\ u=u_0},
\end{equation}
then the linear operator $L_{10}$ should have a discrete zero
eigenvalue. This is a necessary condition for bifurcations, hence it
can be used to determine where a bifurcation might occur. This
condition is not sufficient though. Indeed, if the function
$F(|U|^2, \x)$ in (\ref{e:U}) does not depend explicitly on a
certain spatial dimension $x_j$, i.e., the GNLS equation (\ref{e:U})
is translation-invariant along the $x_j$-dimension, then for any
solitary wave $u(\x; \mu)$, $L_1u_{x_j}=0$, i.e., $L_1$ has a
discrete zero eigenvalue. But this zero eigenvalue of $L_1$ only
corresponds to a spatial translation of $u(\x; \mu)$ and does not
imply solitary wave bifurcations. More will be said on this issue in
the later text (see Remark 3 in Sec. \ref{sec:classification}).

In the next section, we will derive \emph{sufficient} conditions for
three major types of solitary wave bifurcations. To simplify the
analysis, we will focus on the case where this zero eigenvalue of
$L_{10}$ is simple. Hence we introduce the following assumption.

\textbf{Assumption 1 } \ Suppose at a certain propagation constant
$\mu=\mu_0$, $L_{10}$ has a zero eigenvalue. Then it is assumed that
this zero eigenvalue of $L_{10}$ is simple and discrete.

This assumption is satisfied for almost all one-dimensional
bifurcations and many higher-dimensional bifurcations. The case of
$L_{10}$'s zero eigenvalue being multi-fold (repeated) can be
similarly treated, and that will be done elsewhere.

\textbf{Remark 1} \ Due to Assumption 1, the zero eigenvalue of
$L_{10}$ is simple and discrete, thus this zero eigenvalue is not
embedded inside the continuous spectrum of $L_{10}$. This means that
the solitary wave $u_0(\x)$ at $\mu=\mu_0$ is not an embedded
soliton \cite{Yang_SIAM}. This fact allows us to construct solitary
waves in the vicinity of $\mu=\mu_0$ by perturbation series
expansions without worrying about continuous-wave tails beyond all
orders of the perturbation expansion
\cite{Yang_SIAM,Pomeau_1988,Grimshaw_1995,Calvo_Akylas_1997}.

Under Assumption 1, we denote the single discrete (localized)
eigenfunction of $L_{10}$ at the zero eigenvalue as $\psi(\x)$,
i.e.,
\begin{equation}
L_{10}\psi=0.
\end{equation}
Since $L_{10}$ is a real operator, the eigenfunction $\psi$ can be
normalized to be a real function. Thus $\psi$ will be taken as a
real function in the remainder of this article. We also denote
\begin{equation} \label{n:G}
G(u;\x)=F(u^2;\x)u, \quad G_k(\x)=\partial_u^k G|_{u=u_0}, \ k=0, 1,
2, 3, \dots.
\end{equation}
These notations will be used in the next sections.

\section{The main results}
\label{sec:classification}

In this section, we present sufficient analytical conditions for
three major types of solitary wave bifurcations, namely, the
saddle-node bifurcation, the pitchfork bifurcation, and the
transcritical bifurcation. In addition, power diagrams of these
solitary waves near bifurcation points will also be described.

First we explain what these three bifurcations are. A saddle-node
bifurcation is where on one side of the bifurcation point $\mu_0$,
there are no solitary wave solutions; but on the other side of
$\mu_0$, there are two distinct solitary wave branches. These two
branches merge with each other as $\mu\to \mu_0$. This bifurcation
is also called a fold bifurcation in the literature (following a
similar practice in dynamical systems \cite{Murdock}). Examples of
this bifurcation in the GNLS equation (\ref{e:u}) can be found in
\cite{Yang_SIAM,Panos_2005,Kapitula_2006,Akylas_2012}. A pitchfork
bifurcation is where on one side of the bifurcation point $\mu_0$,
there is a single solitary wave branch; but on the other side of
$\mu_0$, there are three distinct solitary wave branches. One of
these three branches is a smooth continuation of the single solution
branch from the other side of $\mu_0$, but the other two branches
are new and they bifurcate out at the bifurcation point $\mu_0$. As
$\mu\to \mu_0$, these two new solution branches merge with the
smooth branch. Examples of pitchfork bifurcations reported so far
are all symmetry-breaking bifurcations
\cite{Akylas_2012,Weinstein_2004,Panos_2005_pitchfork,Weinstein_2008,Sacchetti_2009,Panos_2009,Kirr_2011,Malomed_pitchfork},
where a smooth branch of symmetric or antisymmetric solitary waves
exists on both sides of the bifurcation point, but two new branches
of asymmetric solutions appear on only one side of the bifurcation
point. A transcritical bifurcation is where there are two smooth
branches of solitary waves which exist on both sides of the
bifurcation point $\mu_0$, and these solutions on both branches
approach each other as $\mu\to \mu_0$. So far, no examples of
transcritical bifurcations of solitary waves have been reported in
the GNLS equation (\ref{e:u}) yet (to the author's best knowledge).
But these transcritical bifurcations do exist in Eq. (\ref{e:u}),
and one such example will be presented in Sec. \ref{sec:examples} of
this article.

The main result of this article is the following theorem which gives
sufficient analytical conditions for the above three major types of
solitary wave bifurcations.

\vspace{0.3cm} \textbf{Theorem 1} \ Under Assumption 1, the
following three statements hold.
\begin{enumerate}
\item If $\langle u_0, \psi\rangle \ne 0$ and $\langle G_2,
\psi^3\rangle \ne 0$, then a saddle-node bifurcation occurs at
$\mu=\mu_0$. When these two non-zero quantities have the same
(opposite) sign, the solutions bifurcate to the right (left) side of
$\mu=\mu_0$.

\item If $\langle u_0, \psi\rangle = \langle G_2,
\psi^3\rangle = 0$, $\langle 1-G_2L_{10}^{-1}u_0, \psi^2\rangle\ne
0$, and $\langle G_3, \psi^4\rangle- 3\langle G_2\psi^2,
L_{10}^{-1}(G_2\psi^2)\rangle\ne 0$, then a pitchfork bifurcation
occurs at $\mu=\mu_0$. When these two non-zero quantities have the
same (opposite) sign, the new solution branches bifurcate to the
right (left) side of $\mu=\mu_0$.

\item If $\langle u_0, \psi\rangle =0$, $\langle G_2,
\psi^3\rangle \ne 0$, and
\[
\langle 1-G_2L_{10}^{-1}u_0,, \psi^2\rangle^2 > \langle G_2,
\psi^3\rangle \langle G_2 (L_{10}^{-1}u_0)^2-2L_{10}^{-1}u_0,
\psi\rangle,
\]
then a transcritical  bifurcation occurs at $\mu=\mu_0$.

\end{enumerate}

It is noted that under the conditions of cases 2 and 3 in this
theorem, real quantities $L_{10}^{-1}u_0$ and
$L_{10}^{-1}(G_2\psi^2)$, which appear in these conditions, exist
(see Lemma 1 in the next section).

Theorem 1 shows that in the generic case of $\langle u_0,
\psi\rangle \ne 0$ and $\langle G_2, \psi^3\rangle \ne 0$, a
saddle-node bifurcation occurs. Pitchfork and transcritical
bifurcations would arise only in more restrictive situations. For
instance, pitchfork bifurcations generally occur only in symmetric
potentials, see
\cite{Akylas_2012,Weinstein_2004,Panos_2005_pitchfork,Weinstein_2008,Sacchetti_2009,Panos_2009,Kirr_2011,Malomed_pitchfork}
and Remark 2 below. Transcritical bifurcations are more rare, which
explains why they have not been found in Eq. (\ref{e:u}) before. The
above situation closely resembles that in finite-dimensional
dynamical systems \cite{GH}. More will be said on this in the end of
Sec. \ref{sec:examples}.

\textbf{Remark 2} \ An important (dominant) class of pitchfork
bifurcations is the symmetry-breaking bifurcation. Suppose the
potential in Eq. (\ref{e:U}) is symmetric, i.e.,
\begin{equation}
F(u^2; -\x)=F(u^2; \x).
\end{equation}
In addition, suppose the solitary wave $u_0(\x)$ has certain
symmetry (even or odd in $\x$), and the eigenfunction $\psi(\x)$ has
the opposite symmetry of $u_0(\x)$ (odd or even), i.e.,
\begin{equation}
u_0(-\x)=\pm u_0(\x), \quad \psi(-\x)=\mp \psi(\x).
\end{equation}
From the notation (\ref{n:G}), we see that
\[ G_2=\left[6u\partial_{u^2}F(u^2; \x)+4u^3\partial^2_{u^2}F(u^2; \x)\right]_{u=u_0}, \]
which has the same symmetry as $u_0(\x)$. Then obviously,
\[
\langle u_0, \psi\rangle = \langle G_2, \psi^3\rangle = 0,
\]
thus the conditions of Case 2 in Theorem 1 are generically
satisfied. Consequently, a pitchfork bifurcation occurs at
$\mu=\mu_0$. In this case, the two bifurcated solutions $u^\pm(\x;
\mu)$ are simply related as
\begin{equation} \label{e:symmetry_breakingupm}
u^-(\x; \mu)=u^+(-\x; \mu).
\end{equation}
In addition, these bifurcated solutions break the symmetry of the
original $u_0(\x)$ solution and are asymmetric, as can be seen from
their asymptotic solution formulae in Eq. (\ref{s:upmcase2}) later.
This explains why this pitchfork bifurcation is often called
symmetry-breaking bifurcation in the literature. To the author's
knowledge, all pitchfork bifurcations reported so far are
symmetry-breaking bifurcations.

\textbf{Remark 3} \ Suppose Eq. (\ref{e:U}) is translation invariant
along a certain spatial dimension $x_j$, i.e., $F(|U|^2, \x)$ in
(\ref{e:U}) does not depend explicitly on $x_j$. If this equation
admits a solitary wave $u_0(\x)$ at $\mu=\mu_0$, then by
differentiating Eq. (\ref{e:u}) with respect to $x_j$, we find that
$L_{10}u_{0, x_j}=0$, thus zero is a discrete eigenvalue of $L_{10}$
with eigenfunction $\psi=u_{0, x_j}$. In this case, simple
calculations show that
\[\langle u_0, \psi\rangle = \langle G_2, \psi^3\rangle =\langle
1-G_2L_{10}^{-1}u_0, \psi^2\rangle=0,
\]
and
\[\langle G_3, \psi^4\rangle- 3\langle G_2\psi^2,
L_{10}^{-1}(G_2\psi^2)\rangle=0.
\]
Thus this case does not fall into any of the three cases in Theorem
1, hence no solitary wave bifurcation can be predicted. This is not
surprising, since a zero eigenvalue induced by translation
invariance does not create solitary wave bifurcations in general.

Power diagrams are important not only for displaying solitary wave
bifurcations but also for predicting stability properties of these
solitary waves \cite{Yang_SIAM}. The power diagrams near these three
types of bifurcations are given in the following theorem.

\textbf{Theorem 2} \ Suppose Assumption 1 holds. Denoting the power
of the solitary wave at the bifurcation point as $P_0=\langle u_0,
u_0\rangle$, then
\begin{enumerate}
\item near the saddle-node bifurcation in Case 1 of Theorem 1,
power functions of the two solution branches $u^\pm(\x; \mu)$ are
\begin{equation}  \label{f:powercase1}
P^\pm(\mu)=P_0 \pm P_1 \cdot (\mu-\mu_0)^{1/2} +O(\mu-\mu_0),
\end{equation}
where the constant $P_1$ is given by
\begin{equation}
P_1=2 \hspace{0.05cm} \langle u_0, \psi\rangle \,
\sqrt{\frac{2\langle u_0, \psi\rangle}{\langle G_2, \psi^3\rangle}}
\  ;
\end{equation}

\item near the pitchfork bifurcation in Case 2 of Theorem 1, the
power function for the smooth solution branch $u^0(\x; \mu)$ is
\begin{equation}  \label{f:powercase2a}
P^0(\mu)=P_0 +P_1^0 \cdot (\mu-\mu_0)+O\{(\mu-\mu_0)^2\},
\end{equation}
where the constant $P_1^0$ is given by
\begin{equation}
P_1^0=2 \hspace{0.03cm} \langle u_0, L_{10}^{-1}u_0 \rangle;
\end{equation}
power functions for the two bifurcated solution branches $u^\pm(\x;
\mu)$ are
\begin{equation} \label{f:powercase2b}
P^\pm(\mu)=P_0+P_1 \cdot (\mu-\mu_0)+O\{(\mu-\mu_0)^{3/2}\},
\end{equation}
where the constant $P_1$ is given by
\begin{equation} \label{f:P1case2}
P_1=2 \hspace{0.03cm} \langle u_0,
L_{10}^{-1}u_0\rangle+\frac{6\hspace{0.03cm} \langle 1-G_2
L_{10}^{-1}u_0, \psi^2\rangle^2}{\langle G_3, \psi^4\rangle-
3\langle G_2\psi^2, L_{10}^{-1}(G_2\psi^2)\rangle} \hspace{0.04cm} ;
\end{equation}

\item near the transcritical bifurcation in Case 3 of Theorem 1,
power functions for the two solution branches are
\begin{equation} \label{f:powercase3}
P^\pm(\mu)=P_0 +P_1 \cdot (\mu-\mu_0)+P_2^\pm \cdot
(\mu-\mu_0)^2+O\{(\mu-\mu_0)^3\},
\end{equation}
where the constants $P_1$ and $P_2^\pm$ are given by
\begin{equation}
P_1=2\langle u_0, L_{10}^{-1}u_0\rangle , \quad P_2^\pm = 2\langle
u_0, \widehat{u}_2^\pm \rangle+\langle u_1^\pm, u_1^\pm\rangle,
\end{equation}
with $u_1^\pm$ specified in Eq. (\ref{s:u1case3}),
$\widehat{u}_2^\pm$ being particular solutions to Eq.
(\ref{e:u2case3}), and $b_1$ in (\ref{s:u1case3})-(\ref{e:u2case3})
given in (\ref{s:c1case3}).
\end{enumerate}

This theorem shows that the power diagram near a saddle-node
bifurcation is a horizontally oriented parabola. Near a pitchfork
bifurcation, power curves of the three solution branches are all
linear functions of $\mu$. In addition, the two bifurcated solution
branches $u^\pm(\x; \mu)$ have the same power slope at the
bifurcation point. In fact, in the dominant case of
symmetry-breaking bifurcations discussed in Remark 2, power curves
$P^\pm(\mu)$ of the two solution branches $u^\pm(\x; \mu)$ are
identical for all $\mu$ both near and not near the bifurcation
point, i.e., $P^+(\mu) \equiv P^-(\mu)$, due to the relation
(\ref{e:symmetry_breakingupm}). It is also important to notice that
the smooth solution branch $u^0(\x; \mu)$ and the bifurcated
solution branches $u^\pm(\x; \mu)$ have different power slopes at
the bifurcation point, i.e., $P_1^0\ne P_1$, because the numerator
in the second term of Eq. (\ref{f:P1case2}) is nonzero for pitchfork
bifurcations (see Theorem 1). Thus, the power diagram near a
pitchfork bifurcation looks like a slanted letter `y'. Near a
transcritical bifurcation, power slopes of the two solution branches
at the bifurcation point are the same, but their curvatures are
different in the generic case. Thus the power diagram near a
transcritical bifurcation comprises two smooth curves tangentially
connected at the bifurcation point. These features of the power
diagrams (for pitchfork and transcritical bifurcations) differ
significantly from their familiar solution-bifurcation diagrams, and
these differences are illustrated schematically in Fig. 1.

The upper row of this figure plots the deviation values $u(\x_0;
\mu)-u_0(\x_0)$, as a function of $\mu$, between solitary waves
$u(\x; \mu)$ away from the bifurcation point and the solitary wave
$u_0(\x)$ at the bifurcation point at a representative (fixed)
$\x_0$ position. These curves are drawn using the leading-order
perturbation series solutions (\ref{s:upm}) (for the saddle-node
bifurcation), (\ref{e:uexpandcase2}) and (\ref{s:upmcase2}) (for the
pitchfork bifurcation), and (\ref{e:uexpandcase3b}) (for the
transcritical bifurcation), which we will derive in the next
section. Notice that these deviation diagrams closely resemble the
corresponding bifurcation diagrams (of the same names) in
finite-dimensional dynamical systems \cite{GH}. The lower row of
Fig. 1 plots the associated power diagrams for the bifurcations in
the upper row. These power curves are drawn using the
power-function's asymptotic formulae (\ref{f:powercase1}),
(\ref{f:powercase2a}), (\ref{f:powercase2b}) and
(\ref{f:powercase3}) in Theorem 2. Notice that the power diagram of
the pitchfork bifurcation has a double-branching structure rather
than the familiar triple-branching structure, and the power diagram
of the transcritical bifurcation has a tangential-intersection
structure rather than the familiar `x'-like crossing structure.
These power-diagram behaviors have no counterparts in
finite-dimensional dynamical systems, and they should be borne in
mind when identifying solitary wave bifurcations in the GNLS
equations (\ref{e:U}).

\begin{figure}[h!]
\centerline{\includegraphics[width=0.7\textwidth]{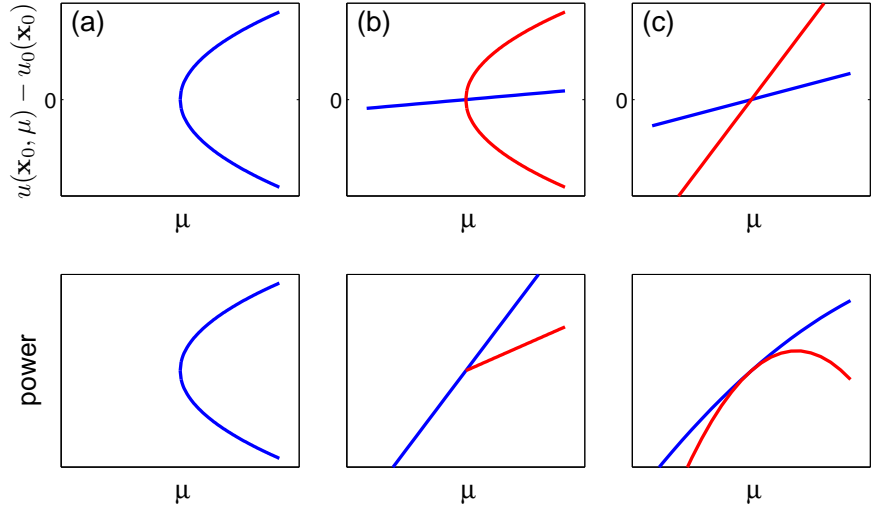}}
\caption{Schematic plots for solitary wave bifurcations (upper row)
and the associated power diagrams (lower row). Column (a):
saddle-node bifurcation; column (b): pitchfork bifurcation; column
(c): transcritical bifurcation. The upper row shows the deviations
$u(\x_0; \mu)-u_0(\x_0)$ versus $\mu$ at a representative $\x_0$
position. These plots are drawn using the perturbation-series
solution (\ref{s:upm}) for (a), (\ref{e:uexpandcase2}) and
(\ref{s:upmcase2}) for (b), and (\ref{e:uexpandcase3b}) for (c). The
power diagrams in the lower row are drawn using the asymptotic
power-function formula (\ref{f:powercase1}) for (a),
(\ref{f:powercase2a}) and (\ref{f:powercase2b}) for (b), and
(\ref{f:powercase3}) for (c). Blue and red colors in columns (b,c)
represent different solution branches.}
\end{figure}

\section{Proofs of the main results}

To prove the main results in Theorems 1 and 2, the following lemma
will be used.

\textbf{Lemma 1 } \ Suppose $f(\x)$ is a localized function. Then
under Assumption 1, the linear inhomogeneous equation
\begin{equation}  \label{e:lemma1}
L_{10}\phi=f
\end{equation}
admits localized solutions $\phi$ if and only if the inhomogeneous
term $f$ is orthogonal to the homogeneous solution $\psi$, i.e.,
\begin{equation}  \label{e:lemma2}
\langle \psi, f\rangle =0.
\end{equation}

This lemma is a direct consequence of the Fredholm Alternative
Theorem. It can also be proved by expanding the localized function
$f(\x)$ and the solution $\phi(\x)$ into the complete set of
eigenfunctions of the Schr\"odinger operator $L_{10}$ and then
solving for $\phi(\x)$ directly.

In the later text, the orthogonality condition (\ref{e:lemma2}) will
be called the solvability condition of the inhomogeneous equation
(\ref{e:lemma1}) (for the existence of localized solutions).

\textbf{Proof of Theorem 1} \ We will use the constructive method to
prove this theorem. Specifically, we will explicitly construct
solitary wave solutions, in the form of perturbation series
expansions, which exist near $\mu=\mu_0$ under the conditions of
this theorem. It will be shown that these perturbation series
solutions can be constructed to all orders. The existence of
available solitary wave solutions near $\mu=\mu_0$ will readily
reveal the type of bifurcations at $\mu=\mu_0$.

\textbf{Case 1: saddle-node bifurcations}

Here we consider the first case of Theorem 1, and show that under
its conditions $\langle u_0, \psi\rangle \ne 0$ and $\langle G_2,
\psi^3\rangle \ne 0$, there exist two solitary-wave branches on only
one side of $\mu=\mu_0$, which merge with each other as $\mu\to
\mu_0$. We will also show that no other solitary wave solutions can
be found near $\mu=\mu_0$. Hence a saddle-node bifurcation occurs
here.

The solitary waves which exist near $\mu=\mu_0$ in this case have
the following perturbation series expansions
\begin{eqnarray}
u(\x; \mu) & = & \sum_{k=0}^\infty (\mu-\mu_0)^{k/2}u_k(\x)
\nonumber   \\
& = & u_0(\x)+(\mu-\mu_0)^{1/2}u_1(\x)+(\mu-\mu_0)u_2(\x) + \cdots.
\label{e:uexpand1}
\end{eqnarray}
Inserting this expansion into Eq. (\ref{e:u}) and recalling the
notations (\ref{n:G}), we get the following sequence of equations
for $u_k$ at order $(\mu-\mu_0)^{k/2}$, $k=0, 1, 2, 3, \dots$:
\begin{eqnarray}
&&\nabla^2u_0-\mu_0 u_0+F(u_0^2, \x)u_0=0,    \label{e:u0} \\
&& L_{10}u_1=0,    \label{e:u1}  \\
&& L_{10}u_2=u_0-\frac{1}{2!}G_2u_1^2,   \label{e:u2} \\
&& L_{10}u_3=u_1-G_2u_1u_2-\frac{1}{3!}G_3u_1^3,   \label{e:u3}  \\
&& L_{10}u_4=u_2-\frac{1}{2!}G_2\left(u_2^2+2u_1u_3\right)-\frac{1}{2!}G_3u_1^2u_2-\frac{1}{4!}G_4u_1^4,   \label{e:u4}  \\
&& \cdots   \nonumber
\end{eqnarray}
The equation (\ref{e:u0}) for $u_0$ is satisfied automatically since
$u_0$ is a solitary wave of Eq. (\ref{e:u}) at $\mu=\mu_0$. The
$u_1$ solution to Eq. (\ref{e:u1}), under Assumption 1, is
\begin{equation}  \label{s:u1}
u_1=b_1\psi,
\end{equation}
where $b_1$ is a constant. The $u_2$ function satisfies the linear
inhomogeneous equation (\ref{e:u2}). Due to Lemma 1, Eq.
(\ref{e:u2}) admits a localized solution for $u_2$ if and only if
\begin{equation}  \label{e:orth1}
\langle \psi, u_0-\frac{1}{2}G_2u_1^2 \rangle=0.
\end{equation}
Inserting the $u_1$ solution (\ref{s:u1}) into this orthogonality
condition and recalling the assumptions of Case 1, we find that
\begin{equation}  \label{e:c12}
b_1=\pm \eta, \quad  \eta \equiv  \sqrt{\frac{2\langle u_0,
\psi\rangle}{\langle G_2, \psi^3\rangle}}.
\end{equation}
Thus, we get two $b_1$ solutions $\pm \eta$ which are opposite of
each other. Inserting the corresponding $u_1$ solutions (\ref{s:u1})
into (\ref{e:uexpand1}), we then get two perturbation series
solutions of $u(\x; \mu)$ as
\begin{equation}  \label{s:upm}
u^\pm (\x; \mu)=u_0(\x)\pm \eta (\mu-\mu_0)^{1/2}
\psi(\x)+O(\mu-\mu_0).
\end{equation}
If $\langle u_0, \psi\rangle$ and $\langle G_2, \psi^3\rangle$ have
the same sign, then $\eta$ is real. Recalling that $u_0(\x)$ and
$\psi(\x)$ are both real localized functions, we see that these two
perturbation series solutions (\ref{s:upm}) give two real-valued
(legitimate) solitary waves when $\mu>\mu_0$, but not when
$\mu<\mu_0$. On the other hand, if $\langle u_0, \psi\rangle$ and
$\langle G_2, \psi^3\rangle$ have the opposite sign, $\eta$ is
purely imaginary. In this case, the perturbation series solutions
(\ref{s:upm}) give two real-valued solitary waves when $\mu<\mu_0$,
but not when $\mu>\mu_0$.

Next we show that the two real localized perturbation series
solutions (\ref{s:upm}), which exist on only one side of
$\mu=\mu_0$, can be constructed to all orders of
$(\mu-\mu_0)^{1/2}$. Let us first consider the $u_2$ equation
(\ref{e:u2}). When $b_1$ is selected from Eq. (\ref{e:c12}), the
orthogonality condition (\ref{e:orth1}) is satisfied. Thus by Lemma
1, localized solutions for $u_2$ exist. Since the inhomogeneous term
and the linear operator $L_{10}$ of (\ref{e:u2}) are both real,
these localized $u_2$ solutions can also be made real. Let us denote
one of such real localized $u_2$ solutions as $\widehat{u}_2$, i.e.,
\[
\widehat{u}_2=L_{10}^{-1}\left(u_0-\frac{1}{2}\eta^2G_2\psi^2\right),
\]
then since $\psi$ is a homogeneous localized solution of
(\ref{e:u2}), the general localized solution of (\ref{e:u2}) is
\begin{equation}  \label{s:u2}
u_2=\widehat{u}_2+b_2\psi,
\end{equation}
where $b_2$ is a constant to be determined.

Now we proceed to the $u_3$ equation (\ref{e:u3}). Inserting the
$u_1$ and $u_2$ solutions (\ref{s:u1}) and (\ref{s:u2}) into
(\ref{e:u3}), we get
\begin{equation} \label{e:u3b}
L_{10}u_3=b_1\left(\psi-G_2\widehat{u}_2\psi-\frac{1}{3!}b_1^2G_3\psi^3-b_2G_2\psi^2\right).
\end{equation}
By Lemma 1, this equation admits localized $u_3$ solutions if and
only if its right hand side is orthogonal to the homogeneous
solution $\psi$. Imposition of this orthogonality condition yields
the $b_2$ value as
\begin{equation} \label{f:c2}
b_2=\frac{\langle 1-G_2\widehat{u}_2-\eta^2G_3\psi^2/3!,
\hspace{0.06cm} \psi^2\rangle}{\langle G_2, \psi^3\rangle},
\end{equation}
which is a real constant. Notice that with this $b_2$ value, the
solution $u_2(\x)$ in (\ref{s:u2}) is the same for both choices $\pm
\eta$ of $b_1$ in the $u_1$ solution (\ref{s:u1}), thus $u_2(\x)$ is
the same for both branches of the perturbation series solutions
$u^\pm(\x; \mu)$ in (\ref{s:upm}). With the $b_2$ value
(\ref{f:c2}), Eq. (\ref{e:u3b}) admits localized solutions
\begin{equation}
u_3=b_1\left(\widehat{u}_3+b_3\psi \right),
\end{equation}
where $\widehat{u}_3$ is a real-valued localized solution of Eq.
(\ref{e:u3b}) but without the $b_1$ factor on its right hand side,
and $b_3$ is a constant. This $b_3$ will be determined from the
solvability (orthogonality) condition of the $u_4$ equation
(\ref{e:u4}) and can be found to be real. Note that
$\widehat{u}_3(\x)$ and $b_3$ are also the same for both branches of
perturbation series solutions $u^\pm(\x; \mu)$.

Proceeding to higher orders and using the method of induction, we
can readily show that all even terms $u_{2n}$ are of the form
\begin{equation}
u_{2n}=\widehat{u}_{2n}+b_{2n}\psi,  \quad n=1, 2, \dots,
\end{equation}
and all odd terms are of the form
\begin{equation}
u_{2n+1}=b_1\left[\widehat{u}_{2n+1}+b_{2n+1}\psi \right], \quad
n=1, 2, \dots,
\end{equation}
where $\widehat{u}_{2n}(\x)$ and $\widehat{u}_{2n+1}(\x)$ are
certain real localized functions, and $b_{2n}$, $b_{2n+1}$ are
certain unique real constants. In addition, $u_{2n}(\x)$,
$\widehat{u}_{2n+1}(\x)$, and $b_{2n+1}$ are the same for both
branches of perturbation series solutions $u^\pm(\x; \mu)$. Thus, by
denoting $\widetilde{u}_{2n+1}= \widehat{u}_{2n+1}+b_{2n+1}\psi$, we
have
\begin{equation}
u_{2n}^\pm=u_{2n}(\x), \quad u_{2n+1}^\pm =\pm \eta \hspace{0.05cm}
\widetilde{u}_{2n+1}(\x).
\end{equation}
Inserting these $u_{2n}^\pm$ and $u_{2n+1}^\pm$ solutions into
(\ref{e:uexpand1}), we obtain two perturbation series solutions for
$u(\x; \mu)$, to all orders of $(\mu-\mu_0)^{1/2}$, as
\begin{eqnarray}
u^\pm(\x; \mu)& = & u_0(\x)+\sum_{n=1}^\infty (\mu-\mu_0)^n
u_{2n}(\x)  \nonumber \\
&& \pm \eta \hspace{0.05cm} (\mu-\mu_0)^{1/2}\left\{ \psi(\x) +
\sum_{n=1}^\infty (\mu-\mu_0)^n \widetilde{u}_{2n+1}(\x) \right\}.
\end{eqnarray}
These two solutions exist on only one side of $\mu=\mu_0$ and are
real and localized. The side of their existence depends on whether
$\eta$ in (\ref{e:c12}) is real or imaginary. When $\mu\to \mu_0$,
$u^\pm(\x; \mu) \to u_0(\x)$, thus $u^\pm(\x; \mu)$ approach each
other and merge at the bifurcation point.

Lastly, we show that except the above two solitary wave branches
which exist on only one side of the bifurcation point, we can not
find other solitary wave solutions near this bifurcation point. For
instance, if we look for smooth solitary-wave branches which exist
on both sides of $\mu=\mu_0$, then their perturbation expansions
should be
\begin{equation} \label{e:uexpandcase1b}
u(\x; \mu)=u_0(\x)+(\mu-\mu_0)u_1(\x)+(\mu-\mu_0)^2u_2(\x)+\cdots.
\end{equation}
When this expansion is substituted into (\ref{e:u}), the $O(1)$
equation is still (\ref{e:u0}) which is satisfied. At
$O(\mu-\mu_0)$, we get the equation for $u_1$ as
\begin{equation} \label{e:u1case1b}
L_{10}u_1=u_0.
\end{equation}
Under conditions of Case 1, $\langle \psi, u_0 \rangle \ne 0$. Thus
by Lemma 1, Eq. (\ref{e:u1case1b}) can not admit any localized
solution for $u_1$. This means that solitary waves with the
perturbation expansion (\ref{e:uexpandcase1b}) can not exist in this
case. We have also searched solitary waves near $\mu=\mu_0$ in other
perturbation series expansions, and could not find such solutions
either. Thus a saddle-node bifurcation occurs at $\mu=\mu_0$.

\textbf{Case 2: pitchfork bifurcations}

Now we consider the second case of Theorem 1. We will show that
under conditions of this case, a smooth branch of solitary waves
exists on both sides of $\mu=\mu_0$. In addition, two other
solitary-wave branches exist on only one side of $\mu=\mu_0$. As
$\mu\to \mu_0$, all these solution branches approach the same
solitary wave $u_0(\x)$. Thus a pitchfork bifurcation occurs at
$\mu= \mu_0$.

(i) We first construct the smooth branch of solitary waves which
exists on both sides of $\mu=\mu_0$. These solitary waves have the
following perturbation series expansion
\begin{equation} \label{e:uexpandcase2}
u^0(\x; \mu) = \sum_{k=0}^\infty (\mu-\mu_0)^{k}u_k(\x).
\end{equation}
Inserting this expansion into Eq. (\ref{e:u}), we get the following
sequence of equations for $u_k$ at orders $(\mu-\mu_0)^k$, $k=0, 1,
2, 3, \dots$:
\begin{eqnarray}
&&\nabla^2u_0-\mu_0 u_0+F(u_0^2, \x)u_0=0,    \label{e:u0case2} \\
&& L_{10}u_1=u_0,    \label{e:u1case2}  \\
&& L_{10}u_2=u_1-\frac{1}{2!}G_2u_1^2,   \label{e:u2case2} \\
&& L_{10}u_3=u_2-G_2u_1u_2-\frac{1}{3!}G_3u_1^3,   \label{e:u3case2}  \\
&& L_{10}u_4=u_3-\frac{1}{2!}G_2\left(u_2^2+2u_1u_3\right)-\frac{1}{2!}G_3u_1^2u_2-\frac{1}{4!}G_4u_1^4,   \label{e:u4case2}  \\
&& \cdots   \nonumber
\end{eqnarray}
The equation (\ref{e:u0case2}) for $u_0$ is satisfied automatically.
Under conditions of Case 2, $\langle \psi, u_0\rangle=0$. Thus by
Lemma 1, the solvability condition for the $u_1$ equation
(\ref{e:u1case2}) is satisfied, hence this equation admits localized
solutions
\begin{equation} \label{s:u1case2}
u_1=\widehat{u}_1+b_1\psi,
\end{equation}
where
\begin{equation}
\widehat{u}_1=L_{10}^{-1}u_0
\end{equation}
is a real localized particular solution to Eq. (\ref{e:u1case2}),
and $b_1$ is a constant to be determined. Inserting this $u_1$
solution into the $u_2$ equation (\ref{e:u2case2}), we get
\begin{equation}  \label{e:u2case2b}
L_{10}u_2=\widehat{u}_1-\frac{1}{2}G_2\widehat{u}_1^2+b_1
\psi(1-G_2\widehat{u}_1)-\frac{1}{2}b_1^2G_2\psi^2.
\end{equation}
By Lemma 1, the solvability condition of this $u_2$ equation is that
its right hand side be orthogonal to the homogeneous solution
$\psi$. Under conditions of Case 2, $\langle G_2, \psi^3\rangle =
0$. Thus this solvability condition gives
\begin{equation}
b_1 \langle 1-G_2L_{10}^{-1}u_0, \psi^2 \rangle = \langle
\frac{1}{2}G_2\widehat{u}_1^2-\widehat{u}_1, \psi\rangle.
\end{equation}
Since the inner product on the left side of this equation is nonzero
under conditions of Case 2, this equation yields a unique $b_1$
value as
\begin{equation} \label{f:c1case2a}
b_1=\frac{\langle G_2\widehat{u}_1^2/2-\widehat{u}_1,
\psi\rangle}{\langle 1-G_2L_{10}^{-1}u_0, \psi^2 \rangle},
\end{equation}
which is a real constant. Hence a real localized solution for $u_1$
has been obtained.

With the above $b_1$ value, the solvability condition of the $u_2$
equation (\ref{e:u2case2b}) is satisfied. Thus this equation admits
a real localized particular solution $\widehat{u}_2$, and its
general solution is
\begin{equation} \label{s:u2case2}
u_2=\widehat{u}_2+b_2\psi,
\end{equation}
where $b_2$ is another constant to be determined.

Inserting this $u_2$ solution into the $u_3$ equation
(\ref{e:u3case2}), this equation becomes
\begin{equation} \label{e:u3case2b}
L_{10}u_3=b_2\psi(1-G_2u_1)+\widehat{u}_2(1-G_2u_1)-\frac{1}{3!}G_3u_1^3.
\end{equation}
By Lemma 1, the solvability condition of this $u_3$ equation is that
its right hand side be orthogonal to $\psi$. Utilizing the $u_1$
solution (\ref{s:u1case2}) and the conditions of Case 2, we see that
\begin{equation}  \label{einnprodnzerocase2}
\langle \psi(1-G_2u_1), \psi \rangle = \langle 1-G_2L_{10}^{-1}u_0,
\psi^2 \rangle \ne 0.
\end{equation}
Thus the solvability condition of Eq. (\ref{e:u3case2b}) yields a
unique real $b_2$ value,
\[
b_2=-\frac{\langle \widehat{u}_2(1-G_2u_1)-G_3u_1^3/3!,
\psi\rangle}{\langle \psi(1-G_2u_1), \psi \rangle},
\]
hence a real localized $u_2$ solution (\ref{s:u2case2}) is then
obtained. At this $b_2$ value, Eq. (\ref{e:u3case2b}) admits a real
localized particular solution $\widehat{u}_3$, and its general
solution is
\begin{equation} \label{s:u3case2}
u_3=\widehat{u}_3+b_3\psi,
\end{equation}
where $b_3$ is another constant to be determined.

Pursuing this calculation to higher orders, it is easy to see that
for any $n\ge 2$, the $u_n$ solution is of the form
\begin{equation} \label{s:uncase2}
u_n=\widehat{u}_n+b_n\psi,
\end{equation}
where $\widehat{u}_n$ is a real localized particular solution of the
$u_n$ equation, and $b_n$ is a constant to be determined from the
solvability condition of the $u_{n+1}$ equation. The $u_{n+1}$
equation is always of the form
\begin{equation}  \label{e:un+1case2}
L_{10}u_{n+1}=(1-G_2u_1)u_n+{\cal F}_{n+1}(u_0, u_1, \dots,
u_{n-1};\x),
\end{equation}
where ${\cal F}_{n+1}$ is some real function which depends on the
already-obtained real localized solutions $u_0, u_1$, $\dots$,
$u_{n-1}$ as well as $\x$. Inserting the $u_n$ solution
(\ref{s:uncase2}) into (\ref{e:un+1case2}) and utilizing Eq.
(\ref{einnprodnzerocase2}), the solvability condition of
(\ref{e:un+1case2}) is met at a unique real $b_n$ value, hence a
real localized $u_n$ solution (\ref{s:uncase2}) is obtained.
Meanwhile, since the solvability condition of (\ref{e:un+1case2}) is
met, a real localized particular solution $\widehat{u}_{n+1}$
exists, and the general $u_{n+1}$ solution is of the form
(\ref{s:uncase2}) with the index $n$ replaced by $n+1$. This process
then repeats itself. Hence a real-valued and localized perturbation
series solution (\ref{e:uexpandcase2}) is constructed to all orders,
and it gives a branch of real-valued solitary waves $u^0(\x; \mu)$
which exists on both sides of $\mu=\mu_0$ and depends smoothly on
$\mu$.

(ii) Next we construct two additional solitary wave branches which
exist on only one side of $\mu=\mu_0$ and merge with the above
smooth solution branch at $\mu=\mu_0$. These additional solitary
wave branches have the following perturbation series expansion
\begin{equation}
u(\x; \mu)  =  \sum_{k=0}^\infty (\mu-\mu_0)^{k/2}u_k(\x).
 \label{e:uexpandcase22}
\end{equation}
This perturbation series is of the same form as (\ref{e:uexpand1})
in Case 1. Thus when this perturbation series is substituted into
Eq. (\ref{e:u}), the resulting equations for $u_k$ are the same as
(\ref{e:u0})-(\ref{e:u4}) before. But since the conditions for Case
2 are different from those for Case 1, solutions $u_k$ for the
perturbation series here will differ from those in
(\ref{e:uexpand1}), as we will show below.

First, the equation (\ref{e:u0}) for $u_0$ is satisfied
automatically since $u_0$ is a solitary wave of (\ref{e:u}) at
$\mu=\mu_0$. The solution $u_1$ to Eq. (\ref{e:u1}), under
Assumption 1, is
\begin{equation}  \label{s:u1case22}
u_1=b_1\psi,
\end{equation}
where $b_1$ is a constant to be determined. Inserting this $u_1$
solution into the $u_2$ equation (\ref{e:u2}), we get
\begin{equation}  \label{e:u2case22}
L_{10}u_2=u_0-\frac{1}{2}b_1^2G_2\psi^2.
\end{equation}
Due to conditions of Case 2, both $u_0$ and $G_2\psi^2$ are
orthogonal to $\psi$. Thus by Lemma 1, both $L_{10}^{-1}u_0$ and
$L_{10}^{-1}(G_2\psi^2)$ exist and are certain real localized
functions. Hence the solution $u_2$ to Eq. (\ref{e:u2case22}) is
\begin{equation}  \label{s:u2case22}
u_2=L_{10}^{-1}u_0-\frac{1}{2}b_1^2 L_{10}^{-1}(G_2\psi^2)+b_2\psi,
\end{equation}
where $b_2$ is another constant to be determined. Inserting these
$u_1$ and $u_2$ solutions into (\ref{e:u3}), the $u_3$ equation is
\begin{equation}  \label{e:u3case22}
L_{10}u_3=b_1\left\{(1-G_2L_{10}^{-1}u_0)\psi-\frac{1}{3!}b_1^2\left[G_3\psi^3-3
G_2\psi L_{10}^{-1}(G_2\psi^2)\right]-b_2G_2\psi^2\right\}.
\end{equation}
In view of the conditions of Case 2, the solvability condition of
this $u_3$ equation yields the $b_1$ value as
\begin{equation} \label{e:c12case2}
b_1=\pm \nu, \quad \nu \equiv \sqrt{\frac{6 \hspace{0.03cm} \langle
1-G_2L_{10}^{-1}u_0, \psi^2\rangle}{\langle G_3, \psi^4\rangle-
3\langle G_2\psi^2, L_{10}^{-1}(G_2\psi^2)\rangle}} \ .
\end{equation}
Two $b_1$ values $\pm \nu$ are obtained which are opposite of each
other. Inserting the corresponding $u_1$ solutions
(\ref{s:u1case22}) into (\ref{e:uexpandcase22}), we then get two
solutions $u^\pm (\x; \mu)$ as perturbation series
\begin{equation}  \label{s:upmcase2}
u^\pm (\x; \mu)=u_0(\x)\pm \nu (\mu-\mu_0)^{1/2}
\psi(\x)+O(\mu-\mu_0),
\end{equation}
where $\nu$ is given in (\ref{e:c12case2}). If the numerator and
denominator under the square root of (\ref{e:c12case2}) have the
same sign, then $\nu$ is real. In this case, two real localized
perturbation series solutions (\ref{s:upmcase2}) are obtained when
$\mu>\mu_0$. If the numerator and denominator in (\ref{e:c12case2})
have the opposite sign, then $\nu$ is purely imaginary. In this
case, two real localized perturbation series solutions
(\ref{s:upmcase2}) are obtained when $\mu<\mu_0$.

Next we show that the two real localized perturbation series
solutions (\ref{s:upmcase2}), which exist on only one side of
$\mu=\mu_0$, can be constructed to all orders of
$(\mu-\mu_0)^{1/2}$. With the choice of $b_1$ values in
(\ref{e:c12case2}), the solvability condition of the $u_3$ equation
(\ref{e:u3case22}) is met, thus the $u_3$ solution is
\begin{equation}
u_3=b_1\left[\widehat{u}_3-b_2L_{10}^{-1}(G_2\psi^2)+b_3\psi\right],
\end{equation}
where $\widehat{u}_3$ is a real localized function which satisfies
the equation
\begin{equation}  \label{e:u3hatcase22}
L_{10}\widehat{u}_3=(1-G_2L_{10}^{-1}u_0)\psi
-\frac{1}{3!}\nu^2\left[G_3\psi^3-3 G_2\psi
L_{10}^{-1}(G_2\psi^2)\right],
\end{equation}
and $b_3$ is a constant to be determined. Inserting these $u_1, u_2$
and $u_3$ solutions into (\ref{e:u4}), the $u_4$ equation becomes
\begin{eqnarray}  \label{e:u4case22}
L_{10}u_4 & = & b_2\left\{(1-G_2L_{10}^{-1}u_0)\psi-\frac{1}{2}
b_1^2\left[G_3\psi^3-3 G_2\psi L_{10}^{-1}(G_2\psi^2)\right]\right\}
\nonumber \\
&& -\frac{1}{2}b_2^2G_2\psi^2+{\cal F}_4(u_0, \psi, b_1^2,
\x)-b_1^2b_3G_2\psi^2,
\end{eqnarray}
where ${\cal F}_4$ is a real localized function which depends on
$u_0$, $\psi$, $b_1^2$ and other already-obtained real functions
(such as $\widehat{u}_3$). Utilizing the $b_1$ formula
(\ref{e:c12case2}) as well as conditions of Case 2, the solvability
condition of this $u_4$ equation is met at the unique real $b_2$
value,
\begin{equation}  \label{e:c2case2}
b_2=\frac{\langle {\cal F}_4, \psi \rangle}{2\langle
1-G_2L_{10}^{-1}u_0, \psi^2\rangle}.
\end{equation}
When this $b_2$ value is inserted into (\ref{s:u2case22}), a real
localized $u_2$ solution is then obtained. Notice that this $b_2$ is
the same for both choices $\pm \nu$ of $b_1$ in the $u_1$ solution
(\ref{s:u1case22}), thus $u_2(\x)$ is the same for both branches of
the perturbation series solutions $u^\pm(\x; \mu)$ in
(\ref{s:upmcase2}).

For the $b_2$ value given in (\ref{e:c2case2}), the solvability
condition of the $u_4$ equation (\ref{e:u4case22}) is satisfied,
thus this equation admits the following localized solution
\begin{equation}
u_4=\widehat{u}_4-\nu^2b_3L_{10}^{-1}(G_2\psi^2)+b_4\psi,
\end{equation}
where $\widehat{u}_4$ is a real localized function which satisfies
the $u_4$ equation (\ref{e:u4case22}) but without the last ($b_3$)
term, and $b_4$ is a new constant to be determined.

Starting from $n\ge 5$, the $u_n$ equation can be derived from
(\ref{e:u}) and the expansion (\ref{e:uexpandcase22}), and is all of
the form
\begin{equation}
L_{10}u_{2n+1}=u_{2n-1}-G_2(u_1u_{2n}+u_2u_{2n-1})-\frac{1}{2}G_3u_1^2u_{2n-1}+b_1{\cal
H}_{2n+1}(u_0, \psi, b_1^2, \x), \quad n\ge 2,
\end{equation}
\begin{equation}
L_{10}u_{2n+2}=u_{2n}-G_2(u_1u_{2n+1}+u_2u_{2n})-\frac{1}{2}G_3u_1^2u_{2n}+{\cal
H}_{2n+2}(u_0, \psi, b_1^2, \x), \quad n\ge 2,
\end{equation}
where ${\cal H}_{2n+1}$ and ${\cal H}_{2n+2}$ are real localized
functions which depend on $u_0$, $\psi$, $b_1^2$ and other already
fully determined real quantities (such as $\widehat{u}_3$,
$\widehat{u}_4$, $b_2$, etc). Using the method of induction as well
as conditions of Case 2, we can show that all $u_n$ solutions are of
the form
\begin{equation}  \label{s:u2n+1case2}
u_{2n+1}=b_1\left[\widehat{u}_{2n+1}-b_{2n}L_{10}^{-1}(G_2\psi^2)+b_{2n+1}\psi\right],
\quad n\ge 1,
\end{equation}
\begin{equation}  \label{s:u2n+2case2}
u_{2n+2}=\widehat{u}_{2n+2}-\nu^2b_{2n+1}L_{10}^{-1}(G_2\psi^2)+b_{2n+2}\psi,
\quad n\ge 1,
\end{equation}
where $\widehat{u}_{2n+1}$ and $\widehat{u}_{2n+2}$ are certain real
localized functions, and $b_{2n+1}$, $b_{2n+2}$ are real constants
which are determined uniquely from the solvability conditions of the
$u_{2n+3}$ and $u_{2n+4}$ equations. We can also show that
$\widehat{u}_{2n+1}$, $\widehat{u}_{2n+2}$, $b_{2n+1}$ and
$b_{2n+2}$ depend on $b_1^2$ as a whole and are thus the same for
both solution branches $u^{\pm}(\x; \mu)$. Inserting these solutions
into the perturbation series (\ref{e:uexpandcase22}), we obtain two
branches of solitary waves $u^{\pm}(\x; \mu)$ whose perturbation
series expansions are
\begin{eqnarray} \label{e:upmexpansioncase2}
u^\pm(\x; \mu)& = & u_0(\x)+\sum_{n=1}^\infty (\mu-\mu_0)^n
u_{2n}(\x)  \nonumber \\
&&  \pm \nu (\mu-\mu_0)^{1/2} \left\{ \psi(\x) + \sum_{n=1}^\infty
(\mu-\mu_0)^n \widetilde{u}_{2n+1}(\x) \right\},
\end{eqnarray}
where real localized functions $u_{2n}$ are given by
(\ref{s:u2case22}) and (\ref{s:u2n+2case2}), real localized
functions $\widetilde{u}_{2n+1}$ are as $u_{2n+1}$ in
(\ref{s:u2n+1case2}) but without the $b_1$ factor, and $\nu$ is
given in (\ref{e:c12case2}). These two real solitary waves exist on
the side of $\mu>\mu_0$ ($\mu<\mu_0$) when $\nu$ is real (purely
imaginary). When $\mu\to \mu_0$, they both approach $u_0(\x)$, thus
these $u^\pm(\x; \mu)$ solution branches merge with the smooth
$u^0(\x; \mu)$ solution branch in (\ref{e:uexpandcase2}) at
$\mu=\mu_0$.

The existence of the smooth solution branch $u^0(\x; \mu)$ in
(\ref{e:uexpandcase2}) on both sides of $\mu=\mu_0$ as well as two
additional solution branches $u^\pm(\x; \mu)$ in
(\ref{e:upmexpansioncase2}) on only one side of $\mu=\mu_0$
indicates that a pitchfork bifurcation occurs at $\mu=\mu_0$.

\textbf{Case 3: transcritical bifurcations}

Now we consider the third case of Theorem 1. We will show that under
conditions of this case, two smooth branches of solitary waves exist
on both sides of $\mu=\mu_0$, and these branches intersect at
$\mu=\mu_0$ where solitary waves on the two branches become
identical. Thus a transcritical bifurcation occurs at $\mu= \mu_0$.

In this third case, we seek solitary wave solutions which exist on
both sides of $\mu=\mu_0$ and depend smoothly on $\mu$ near
$\mu=\mu_0$. The perturbation series expansion of such solutions is
\begin{equation} \label{e:uexpandcase3}
u(\x; \mu) = \sum_{k=0}^\infty (\mu-\mu_0)^{k}u_k(\x).
\end{equation}
The form of this expansion is the same as (\ref{e:uexpandcase2}) in
Case 2, thus the equations for $u_k$ are also the same as
(\ref{e:u0case2})-(\ref{e:u4case2}) before. However, the solutions
to these equations will differ from the previous ones in Case 2 due
to different conditions of the present case.

The $u_0$ equation (\ref{e:u0case2}) is satisfied automatically
since $u_0$ is a solitary wave of Eq. (\ref{e:u}) at $\mu=\mu_0$.
Under conditions of Case 3, the solvability condition of the $u_1$
equation (\ref{e:u1case2}), $\langle u_0, \psi\rangle=0$, is met.
Thus by Lemma 1, localized $u_1$ solutions of the form
\begin{equation}  \label{s:u1case3}
u_1=L_{10}^{-1}u_0+b_1\psi
\end{equation}
are admitted. Here $L_{10}^{-1}u_0$ is a real and localized
particular solution to Eq. (\ref{e:u1case2}), and $b_1$ is a
constant to be determined. Inserting this $u_1$ solution into the
$u_2$ equation (\ref{e:u2case2}), this equation becomes
\begin{equation} \label{e:u2case3}
L_{10}u_2=L_{10}^{-1}u_0-\frac{1}{2}G_2(L_{10}^{-1}u_0)^2+b_1(1-G_2L_{10}^{-1}u_0)\psi-\frac{1}{2}b_1^2G_2\psi^2.
\end{equation}
The solvability condition of this equation gives the following
quadratic equation for $b_1$:
\begin{equation}  \label{e:c1case3}
\langle G_2, \psi^3\rangle b_1^2-2\langle 1-G_2L_{10}^{-1}u_0,
\psi^2\rangle b_1+\langle G_2(L_{10}^{-1}u_0)^2-2L_{10}^{-1}u_0,
\psi\rangle=0.
\end{equation}
Under conditions of Case 3, the coefficient of the $b_1^2$ term in
this quadratic equation is nonzero, and
\[
\Delta \equiv \langle 1-G_2L_{10}^{-1}u_0,, \psi^2\rangle^2 -\langle
G_2, \psi^3\rangle \langle G_2 (L_{10}^{-1}u_0)^2-2L_{10}^{-1}u_0,
\psi\rangle >0.
\]
Thus this quadratic equation admits the following two real roots
\begin{equation}  \label{s:c1case3}
b_1=b_1^\pm \equiv \frac{\langle 1-G_2L_{10}^{-1}u_0,
\psi^2\rangle\pm \sqrt{\Delta}}{\langle G_2, \psi^3\rangle }.
\end{equation}
For each of these two $b_1$ values, a real localized $u_1$ solution
(\ref{s:u1case3}) is obtained. In addition, a real and localized
particular solution $\widehat{u}_2$ to the $u_2$ equation
(\ref{e:u2case3}) exists, hence the $u_2$ solution is
\begin{equation}  \label{s:u2case3}
u_2=\widehat{u}_2+b_2\psi,
\end{equation}
where $b_2$ is a new constant to be determined.

Inserting the above $u_2$ solution into the $u_3$ equation
(\ref{e:u3case2}), we get
\begin{equation}  \label{e:u3case3b}
L_{10}u_3=b_2(1-G_2u_1)\psi+(1-G_2u_1)\widehat{u}_2-\frac{1}{3!}G_3u_1^3.
\end{equation}
Utilizing the $u_1$ solution (\ref{s:u1case3}) and the $b_1$ formula
(\ref{s:c1case3}), we find that
\begin{equation}  \label{e:nonzeroinnprodcase3}
\langle (1-G_2u_1)\psi, \psi\rangle=\langle 1-G_2L_{10}^{-1}u_0,
\psi^2\rangle -b_1\langle G_2, \psi^3\rangle=\mp \sqrt{\Delta}\ne 0.
\end{equation}
Thus the solvability condition of Eq. (\ref{e:u3case3b}) yields a
real constant $b_2$ as
\[
b_2=-\frac{\langle (1-G_2u_1)\widehat{u}_2-G_3u_1^3/3!,
\hspace{0.06cm} \psi\rangle}{\langle (1-G_2u_1)\psi, \hspace{0.04cm}
\psi\rangle}.
\]
For this $b_2$ value, the solvability condition of the $u_3$
equation (\ref{e:u3case3b}) is satisfied, thus this equaiton admits
a real localized particular solution $\widehat{u}_3$, and the
general $u_3$ solution is
\begin{equation}
u_3=\widehat{u}_3+b_3\psi,
\end{equation}
where $b_3$ is another constant to be determined.

Pursuing this calculation to higher orders, it is easy to see that
for any $n\ge 2$, the $u_n$ solution is of the form
\begin{equation} \label{s:uncase3}
u_n=\widehat{u}_n+b_n\psi,
\end{equation}
where $\widehat{u}_n$ is a real localized particular solution of the
$u_n$ equation, and $b_n$ is a real constant to be determined from
the solvability condition of the $u_{n+1}$ equation. The $u_{n+1}$
equation is always of the form
\begin{equation}  \label{e:un+1case3}
L_{10}u_{n+1}=(1-G_2u_1)u_n+{\cal F}_{n+1}(u_0, u_1, \dots,
u_{n-1};\x),
\end{equation}
where ${\cal F}_{n+1}$ is some real function which depends on the
already-obtained real localized solutions $u_0, u_1$, $\dots$,
$u_{n-1}$ and $\x$. Inserting the $u_n$ solution (\ref{s:uncase3})
into (\ref{e:un+1case3}) and in view of Eq.
(\ref{e:nonzeroinnprodcase3}), the solvability condition of
(\ref{e:un+1case3}) then yields a unique real value for the constant
$b_n$.

In the above solution process, since $b_1$ can take either one of
the two real roots $b_1^\pm$ in (\ref{s:c1case3}), $u_1$ in
(\ref{s:u1case3}) then has two corresponding solutions $u_1^\pm$.
These two $u_1$ solutions cascade up to higher orders, and thus all
$u_n$ functions have two solutions $u_n^\pm$. Consequently, two
real-valued and localized perturbation series solutions
\begin{equation} \label{e:uexpandcase3b}
u^\pm(\x; \mu) = u_0(\x)+\sum_{k=1}^\infty
(\mu-\mu_0)^{k}u_k^\pm(\x)
\end{equation}
are obtained to all orders, and they provide two branches of
real-valued solitary waves $u^\pm(\x; \mu)$ which exist on both
sides of $\mu=\mu_0$ and depend smoothly on $\mu$. When $\mu\to
\mu_0$, both $u^\pm(\x; \mu)$ approach $u_0(\x)$, thus these two
solution branches intersect at $\mu=\mu_0$. As a result, a
transcritical bifurcation occurs at $\mu=\mu_0$. This completes the
proof of Theorem 1. $\Box$

Next, we prove Theorem 2 on power diagrams near bifurcation points.

\textbf{Proof of Theorem 2 } \ The power formula
(\ref{f:powercase1}) of saddle-node bifurcations can be derived
easily from the perturbation series solutions (\ref{e:uexpand1}) and
the $u_1$ solution (\ref{s:u1}) with $b_1$ given by Eq.
(\ref{e:c12}). The power formula (\ref{f:powercase2a}) for the
smooth solution branch $u^0(\x; \mu)$ in a pitchfork bifurcation can
be derived readily from the perturbation series solutions
(\ref{e:uexpandcase2}) and the $u_1$ solution (\ref{s:u1case2}). To
derive the power formula (\ref{f:powercase2b}) for the two
bifurcated solution branches in a pitchfork bifurcation, we
substitute the $u_1$, $u_2$ solutions in (\ref{s:u1case22}) and
(\ref{s:u2case22}) into the expansion (\ref{e:upmexpansioncase2}),
and find that the power function is given by (\ref{f:powercase2b}),
where
\begin{equation}
P_1=2\langle u_0, L_{10}^{-1}u_0\rangle+b_1^2\left[\langle \psi,
\psi\rangle-\langle u_0, L_{10}^{-1}(G_2\psi^2)\rangle\right],
\end{equation}
whose value is the same for both bifurcated branches. Since
$L_{10}^{-1}$ is self-adjoint and $L_{10}^{-1}u_0$ exists (by Lemma
1), this $P_1$ coefficient can then be rewritten as
(\ref{f:P1case2}) (here the $b_1$ formula (\ref{e:c12case2}) is also
used). The power formula (\ref{f:powercase3}) for transcritical
bifurcations can be derived easily from the perturbation series
solutions (\ref{e:uexpandcase3b}) and the $u_1$, $u_2$ solutions
(\ref{s:u1case3}), (\ref{s:u2case3}). $\Box$

\section{Numerical examples of solitary wave bifurcations}
\label{sec:examples}

In this section, we present numerical examples for these three types
of solitary wave bifurcations, and compare them with the analytical
results presented in Theorems 1 and 2. So far, examples of
saddle-node and pitchfork bifurcations of solitary waves have been
reported in the GNLS equations (\ref{e:U}) with various potentials
and nonlinearities
\cite{Yang_SIAM,Panos_2005,Kapitula_2006,Akylas_2012,Weinstein_2004,Panos_2005_pitchfork,Weinstein_2008,Sacchetti_2009,Panos_2009,Kirr_2011,Malomed_pitchfork}.
Here we will present some new examples of saddle-node and pitchfork
bifurcations in the GNLS equations which exhibit interesting novel
features. In addition, we will present the first example of
transcritical bifurcation in these GNLS equations.

\noindent \textbf{Example 1:  combined saddle-node and
double-pitchfork bifurcations}

The first example we choose is the one-dimensional GNLS equation
(\ref{e:U}) with a symmetric double-well potential and cubic-quintic
nonlinearity:
\begin{equation}  \label{e:Uexample}
iU_t+U_{xx}-V(x)U+|U|^2U-\gamma |U|^4U=0,
\end{equation}
where the symmetric double-well potential $V(x)$ is taken of the
form
\begin{equation}  \label{e:potential_example1}
V(x)=-V_0\left[\mbox{sech}^2(x+x_0)+\mbox{sech}^2(x-x_0)\right],
\end{equation}
$V_0>0$ is the potential depth, $2x_0$ is the separation between the
two wells, and $\gamma>0$ is the coefficient of the quintic
nonlinearity. Notice that the cubic and quintic nonlinear terms in
(\ref{e:Uexample}) have the opposite sign, and the quintic term
induces a self-defocusing effect which counters the self-focusing
effect of the cubic term. One may also view this opposing
cubic-quintic nonlinearity as a Taylor-series approximation to the
saturable nonlinearity in photorefractive crystals
\cite{Christodoulides_model}. The parameter values in the above GNLS
model are chosen as
\begin{equation}
V_0=2.8, \quad x_0=1.5, \quad \gamma=0.25.
\end{equation}

Solitary waves in Eq. (\ref{e:Uexample}) are sought of the form
(\ref{e:Usoliton}), where $u(x)$ is a real localized function
satisfying the equation
\begin{equation}  \label{e:uexample}
u_{xx}-\mu u-V(x)u+u^3-\gamma u^5=0.
\end{equation}
When $u(x)$ is infinitesimal, the linear Schr\"odinger operator of
Eq. (\ref{e:uexample}) admits a positive symmetric discrete
eigenfunction at eigenvalue $\mu \approx 1.7896$. This eigenmode is
the ground state of the underlying double-well potential. From this
linear (infinitesimal) eigenmode, a family of positive symmetric
solitary waves bifurcates out. The power curve of this
symmetric-soliton family is shown in Fig. 2 (blue curve in the upper
left panel). We have computed the spectra of the linearization
operator $L_1$ for these solitary waves, and found that their
spectra contain a simple zero eigenvalue at three locations marked
by letters `A,B,C' on the power curve. This is evidenced in the
upper right panel of Fig. 2, where the $L_1$-spectra of solitary
waves at these three locations are displayed. Notice that at
locations `A,B', the second largest eigenvalue of the spectrum is
zero, while at location `C', the largest eigenvalue is zero. At
these three locations, solitary waves $u_0(x)$ and eigenfunctions
$\psi(x)$ of the zero eigenvalue in $L_1$'s spectra are plotted in
the lower row of Fig. 2 (as solid blue and dashed red curves
respectively). Notice that eigenfunctions $\psi(x)$ at points `A,B'
are anti-symmetric, while the eigenfunction at point `C' is
symmetric.

\begin{figure}[h!]
\centerline{\includegraphics[width=0.8\textwidth]{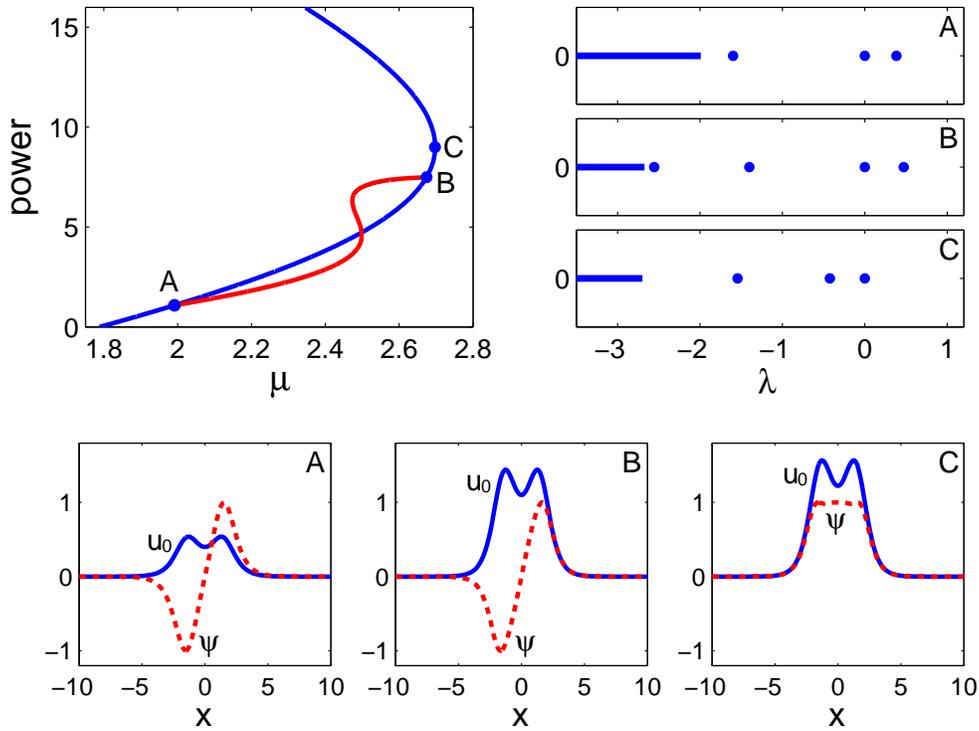}}
\caption{Bifurcations of solitary waves in Example 1. Upper left:
the power diagram; the blue curve is for the family of symmetric
solitary waves, and the red curve is for the family of asymmetric
solitary waves which bifurcate out from points `A, B' through double
pitchfork bifurcations. Upper right: $L_1$'s spectra for solitary
waves at bifurcation points `A,B,C' of the power diagram. Lower row:
solitary waves $u_0(x)$ and eigenfunctions $\psi(x)$ of $L_1$'s zero
eigenvalue at bifurcation points `A,B,C' (the eigenfunctions are
normalized to have unit amplitude). } \label{f:fig2}
\end{figure}

At these `A,B,C' points, zero is a simple discrete eigenvalue of
$L_1$, thus Assumption 1 is met and Theorem 1 applies. In addition,
for the present example,
\[
G_2=6u_0-20\gamma u_0^3, \quad G_3=6-60\gamma u_0^2.
\]
Now we use our analytical criterion in Theorem 1 to determine if and
what bifurcations occur at these points.

At points `A,B', it is easy to see from symmetry that
\[
\langle u_0, \psi\rangle = \langle G_2, \psi^3\rangle = 0.
\]
In addition, when the eigenfunction $\psi$ is normalized to have
unit amplitude (see Fig. 2, lower row), we find numerically that at
point `A',
\[
\langle 1-G_2L_{10}^{-1}u_0, \psi^2\rangle=-4.9313, \quad \langle
G_3, \psi^4\rangle- 3\langle G_2\psi^2,
L_{10}^{-1}(G_2\psi^2)\rangle =-58.4035;
\]
and at point `B',
\[
\langle 1-G_2L_{10}^{-1}u_0, \psi^2\rangle=23.9913, \quad \langle
G_3, \psi^4\rangle- 3\langle G_2\psi^2,
L_{10}^{-1}(G_2\psi^2)\rangle =-110.9244.
\]
Then according to Theorem 1, pitchfork bifurcations occur at both
`A' and `B' points. In addition, the new (asymmetric) solitary waves
bifurcate out on the right side of point `A' and on the left side of
point `B'.

At point `C', we find that
\[
\langle u_0, \psi\rangle=6.4879,   \quad \langle G_2,
\psi^3\rangle=-21.0632,
\]
thus according to Theorem 1, a saddle-node bifurcation occurs at
this point. In addition, the bifurcated solutions appear on the left
side of point `C'.

These analytical predictions of bifurcations prove to be completely
correct. Specifically, at points `A,B', symmetry-breaking pitchfork
bifurcations occur. The two bifurcated asymmetric solitary waves
$u^\pm(x; \mu)$ are related to each other by a mirror reflection in
$x$, i.e., $u^+(-x; \mu)=u^-(x; \mu)$, and their power curves (which
are identical) are displayed as the red line in Fig. 2 (upper left
panel). Notice that these bifurcated solutions appear on the right
side of point `A' and on the left side of point `B', as predicted by
the above analysis. To illustrate solution profiles before and after
these bifurcations, we focus on point `A'. The power diagram near
this bifurcation point is amplified from that in Fig. 2 and shown in
Fig. 3 (first panel from the left). Notice that the power curves
near this bifurcation point are linear functions of $\mu$, in
agreement with Theorem 2 and Fig. 1(b). We have also compared the
slopes of the power curves at point `A' in Fig. 3 with the
analytical power slopes in Eqs.
(\ref{f:powercase2a})-(\ref{f:P1case2}), and found excellent
agreement. At three locations `a,b,c' on the two sides of the
bifurcation point `A' in the power diagram, profiles of the solitary
waves are displayed in Fig. 3(a-c) respectively. Solutions in Fig.
3(a,b) are symmetric and lie on the symmetric branch of the power
diagram (blue line), while the two solutions in Fig. 3(c) are
asymmetric and lie on the asymmetric (bifurcated) branch of the
power diagram (red line). Notice that on the left side of the
bifurcation point, there is a single (symmetric) solitary wave (see
Fig. 3(a)); but on the right side of the bifurcation point, there
are three solitary waves, one symmetric (see Fig. 3(b)) and the
other two asymmetric (see Fig. 3(c)). These behaviors of the
pitchfork bifurcation agree fully with our analytical results as
well as the schematic plots in Fig. 1(b).

\begin{figure}[h!]
\centerline{\includegraphics[width=0.8\textwidth]{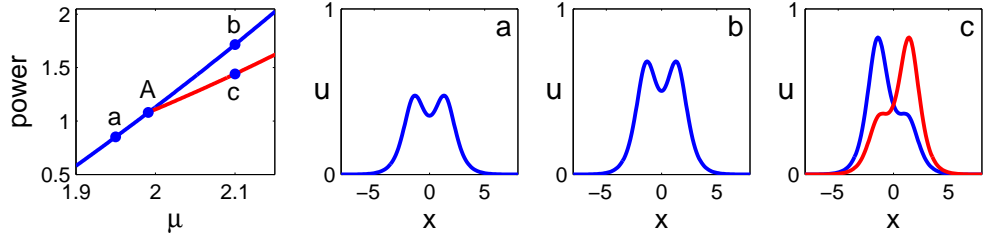}}
\caption{Power diagram and profiles of solitary waves near the
pitchfork bifurcation point `A' in Fig. 2. First panel: power
diagram; (a,b,c) solitary waves at locations marked by the same
letters on the power diagram. }
\end{figure}

It is interesting to observe from Fig. 2 (upper left panel) that the
asymmetric soliton branch starts out from point `A' and terminates
at point `B', thus it appears and then disappears through double
pitchfork bifurcations. In between, its power curve exhibits a `S'
shape, indicating that double saddle-node bifurcations also occur on
this asymmetric branch. These features of bifurcations are quite
novel for the GNLS equations (\ref{e:U}). In a different nonlinear
wave system, namely nonlinear saturable couplers, a similar double
pitchfork bifurcation also exists \cite{Akhmediev_pitchfork_1995}.

At point `C' of Fig. 2, we have found that a saddle-node bifurcation
occurs as predicted. This is already obvious from the power diagram
in Fig. 2, which shows that the power curve turns around at this
point. The power diagram near this saddle-node bifurcation point `C'
is amplified and shown again in Fig. 4. This numerical power curve
is compared with the analytical saddle-node power formula
(\ref{f:powercase1}) and complete agreement is obtained. At two
locations `a,b' of the power curve below and above the bifurcation,
profiles of the solitary waves are displayed in Fig. 4(a,b). These
solutions are all symmetric, and their amplitudes vary while going
through the bifurcation.

\begin{figure}[h!]
\centerline{\includegraphics[width=0.6\textwidth]{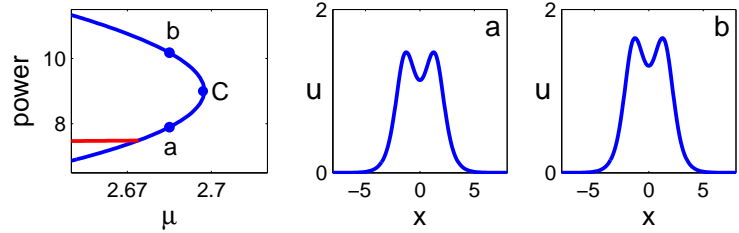}}
\caption{Power diagram and profiles of solitary waves near the
saddle-node bifurcation point `C' in Fig. 2. Left panel: power
diagram; (a,b) solitary waves at locations marked by the same
letters on the power diagram. }
\end{figure}

\noindent \textbf{Example 2: power loop phenomena}

\begin{figure}[h!]
\centerline{\includegraphics[width=0.6\textwidth]{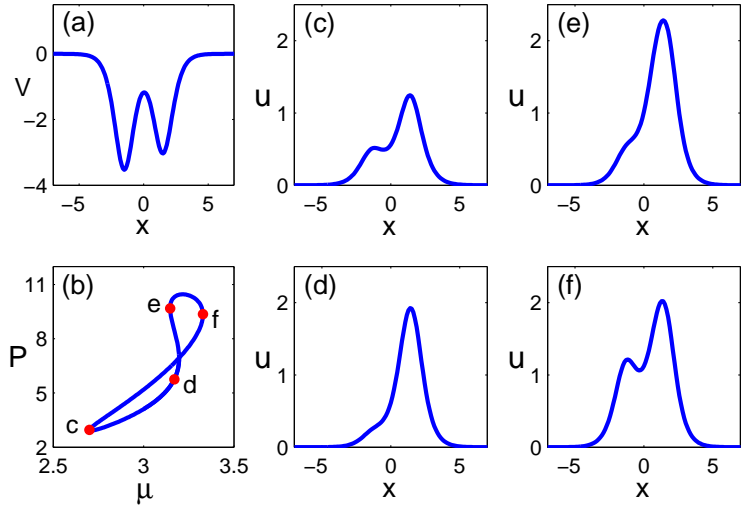}}
\caption{Power-loop phenomenon in Example 2 (i.e., Eq.
(\ref{e:Uexample2})). (a) The asymmetric double-well potential
$V(x)$ in Eq. (\ref{e:Vexample2}); (b) the power loop; (c-f)
profiles of solitary waves at locations marked by the same letters
on the power loop of (b). } \label{f:fig5}
\end{figure}

Our second example is still the GNLS equation with opposing cubic
and quintic nonlinearities,
\begin{equation}  \label{e:Uexample2}
iU_t+U_{xx}-V(x)U+|U|^2U-0.15 |U|^4U=0,
\end{equation}
but the double-well potential $V(x)$ is now asymmetric instead:
\begin{equation} \label{e:Vexample2}
V(x)=-3.5\: \mbox{sech}^2(x+1.5)-3\: \mbox{sech}^2(x-1.5).
\end{equation}
This potential is displayed in Fig. \ref{f:fig5}(a). As usual,
solitary waves in this equation are sought of the form
(\ref{e:Usoliton}), where $u(x)$ is a real localized function. We
find that in this system, there exist a family of positive solitary
waves whose power curve forms a closed loop. This power loop is
displayed in Fig. \ref{f:fig5}(b). This power loop shows that this
family of solitary waves has a non-zero minimal power and a finite
maximal power, and it exists over a finite propagation-constant
interval. In addition, four saddle-node bifurcations are clearly
visible on this loop. We have checked that at these saddle-node
bifurcation points, the bifurcation conditions in Theorem 1 (Case 1)
are all satisfied. At four locations of the power loop, three of
them (`c,e,f') being saddle-node bifurcation points and the
remaining one (`d') slightly below a saddle-node bifurcation point,
profiles of the solitary waves are displayed in Fig.
\ref{f:fig5}(c-f). It is seen that the energy of these solitary
waves resides primarily in the shallower (right) well of the
potential. Thus this family of solitary waves is different from the
family of ground-state solitary waves in this system, whose energy
resides primarily in the deeper (left) well of the potential. One
may notice that this power loop in Fig. \ref{f:fig5}(b) self-crosses
itself in the middle (above point `d'). This power-curve crossing
does not signal a transcritical bifurcation however, because as
$\mu$ approaches this crossing point along the two intersecting
curves, the solitary waves do not approach each other. This power
loop phenomenon of solitary waves has not been reported before in
the GNLS equations (\ref{e:U}) (to the author's knowledge), but a
similar momentum loop phenomenon for solitons sitting on constant
backgrounds has been discovered in the externally driven NLS
equation \cite{momentum_loop}.

\noindent \textbf{Example 3: transcritical bifurcation}

\begin{figure}[h!]
\centerline{\includegraphics[width=0.6\textwidth]{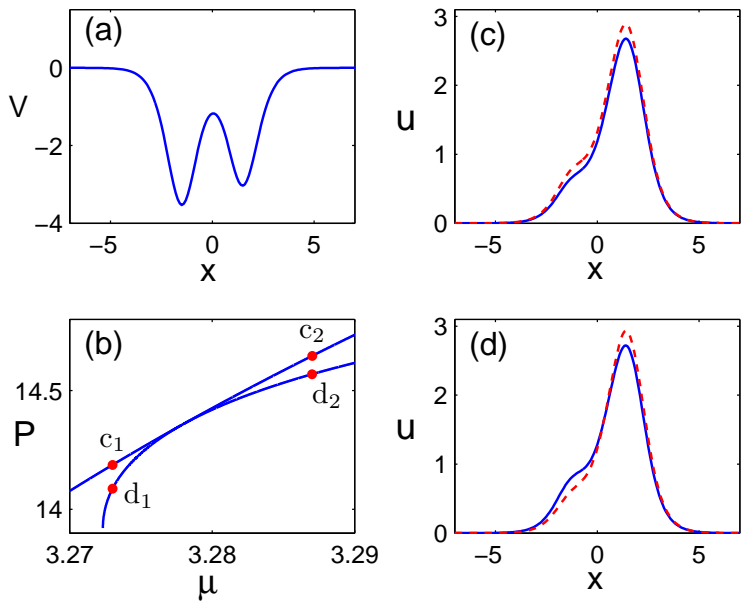}}
\caption{Transcritical bifurcation in Example 3 (see Eq.
(\ref{e:Uexample3})). (a) The asymmetric double-well potential
$V(x)$ in this example; (b) the power diagram; (c) profiles of
solitary waves at locations $c_1$ (solid blue) and $c_2$ (dashed
red) of the upper power curve in (b); (d) profiles of solitary waves
at locations $d_1$ (solid blue) and $d_2$ (dashed red) of the lower
power curve in (b). } \label{f:fig6}
\end{figure}

Our last example is the GNLS equation with competing cubic, quintic
and seventh-power nonlinearities,
\begin{equation}  \label{e:Uexample3}
iU_t+U_{xx}-V(x)U+|U|^2U-0.2 |U|^4U +\kappa |U|^6U=0,
\end{equation}
where $V(x)$ is the same asymmetric double-well potential
(\ref{e:Vexample2}) as in Example 2, and $\kappa$ is a real
constant. In this example, a transcritical bifurcation of solitary
waves is found at
\begin{equation} \label{e:kappac}
\kappa=\kappa_c\approx 0.01247946.
\end{equation}
The power diagram of this bifurcation is shown in Fig.
\ref{f:fig6}(b). We see that two smooth solution branches, namely
the upper $c_1$-$c_2$ branch and the lower $d_1$-$d_2$ branch,
tangentially connect at the bifurcation point $(\mu_0, P_0)\approx
(3.28, 14.35)$. Profiles of solitary waves at the marked $c_1, c_2,
d_1, d_2$ locations on this power diagram are displayed in Fig.
\ref{f:fig6}(c-d). Notice that these solutions are close to each
other since the $c_1, c_2, d_1, d_2$ locations are near the
bifurcation point $(\mu_0, P_0)$. As $\mu$ approaches this
bifurcation point, we find that these solitary waves along both the
lower and upper power branches approach each other, confirming that
this is a transcritical bifurcation. Notice that the power diagram
in Fig. \ref{f:fig6}(b) agrees with the analytical power formula
(\ref{f:powercase3}) of transcritical bifurcations (see also the
schematic power diagram in Fig. 1(c)). In addition, we have checked
the conditions of transcritical bifurcations in Theorem 1 (Case 3),
and found them satisfied here as well.

What would happen to the bifurcations in the above three examples if
the potential or the nonlinearity in those underlying GNLS equations
is slightly perturbed? We have numerically studied this question and
found that in Example 1, when the nonlinearity or the potential is
slightly and arbitrarily perturbed (including perturbations to make
the double-well potential (\ref{e:potential_example1}) asymmetric),
the saddle-node bifurcations (at point `C' of the symmetric-soliton
branch and two others on the asymmetric-soliton branch in Fig.
\ref{f:fig2}) always persist. For the two pitchfork bifurcations in
this example (at points `A,B' of Fig. 2), if the perturbed potential
is still symmetric, then these pitchfork bifurcations would survive;
but if the perturbed potential becomes asymmetric, then these
pitchfork bifurcations are destroyed. In Example 2, the four
saddle-node bifurcations on the power loop of Fig. \ref{f:fig5}
always persist under weak perturbations in the nonlinearity or the
potential. In Example 3, the transcritical bifurcation is extremely
sensitive and is destroyed under generic small perturbations to the
system (such as when $\kappa\ne \kappa_c$). From these numerical
results, we conclude that saddle-node bifurcations are generic and
robust under weak perturbations to the system; pitchfork
bifurcations are generally reliant on a symmetric potential; and
transcritical bifurcations are very fragile and generally disappear
under perturbations. These behaviors are consistent with similar
statements on these bifurcations (based on the bifurcation
conditions) below Theorem 1.

\section{Summary and discussion}

In this paper, we classified solitary wave bifurcations in the
generalized NLS equations (\ref{e:U}) with arbitrary nonlinearities
and external potentials in arbitrary spatial dimensions. Sufficient
analytical conditions were derived for three major types of solitary
wave bifurcations, namely saddle-node bifurcations, pitchfork
bifurcations and transcritical bifurcations. These conditions show
that the generic solitary wave bifurcation is the saddle-node
bifurcation; the pitchfork bifurcation generally requires certain
symmetry conditions; and the transcritical bifurcation is rare. For
these bifurcations, shapes of power diagrams near the bifurcation
points were also obtained. It was shown that the power diagram for a
pitchfork bifurcation exhibits double branching rather than the
familiar triple branching, and the power diagram for a transcritical
bifurcation features two curves tangentially touching each other
rather than the familiar `x'-crossing. Numerical examples for these
three types of bifurcations were presented as well. These examples
show novel features such as power loops and double pitchfork
bifurcations. The example of transcritical bifurcation seems to be
the first report of such bifurcation in the generalized NLS
equations (\ref{e:U}).

The results in this paper are important not only for a general
classification and understanding of solitary wave bifurcations in
the generalized NLS equations (\ref{e:U}). More importantly, the
bifurcation conditions in Theorem 1 will be the basis for a general
treatment of linear stability of solitary waves near bifurcation
points. This stability analysis lies outside the scope of the
present article and will be reported elsewhere. We do want to say
here that the stability properties of solitary waves near
bifurcation points in Eq. (\ref{e:U}) show some important
qualitative differences from those in finite-dimensional dynamical
systems \cite{GH}. Details will be presented in a forthcoming
article.

\section*{Acknowledgment}
This work is supported in part by the Air Force Office of Scientific
Research (Grant USAF 9550-09-1-0228) and the National Science
Foundation (Grant DMS-0908167).


\end{document}